\documentclass[amsmath,amssymb,superscriptaddress,nobalancelastpage,aps,prl,twocolumn, longbibliography]{revtex4-2}

\usepackage{graphicx}
\usepackage{varioref}
\usepackage{xr-hyper}
\usepackage{xcolor}
\usepackage{nicefrac}
\usepackage{xfrac}
\usepackage{hyperref}
\hypersetup{colorlinks,linkcolor=blue, urlcolor=blue,citecolor=blue}
\usepackage{ulem}
\usepackage{lineno}
\usepackage{amsmath} 
\usepackage{amssymb}
\usepackage{siunitx}

\newcommand\purplesout{\bgroup\markoverwith{\textcolor{purple}{\rule[0.5ex]{2pt}{0.4pt}}}\ULon}
\newcommand\tealsout{\bgroup\markoverwith{\textcolor{teal}{\rule[0.5ex]{2pt}{0.4pt}}}\ULon}

\newcommand{\dz}{$d_{z^2}$}
\newcommand{\dx}{$d_{x^2-y^2}$}

\begin{document}



\title{Engineering Phase Competition Between Stripe Order and Superconductivity in \texorpdfstring{La$_{1.88}$Sr$_{0.12}$CuO$_4$}{LSCO}}

\author{J. Küspert}
\email{julia.kuespert@physik.uzh.ch}
\affiliation{Physik-Institut, Universit\"{a}t Z\"{u}rich, Winterthurerstrasse 
190, CH-8057 Z\"{u}rich, Switzerland}

\author{I. Bia\l{}o}
\affiliation{Physik-Institut, Universit\"{a}t Z\"{u}rich, Winterthurerstrasse 
190, CH-8057 Z\"{u}rich, Switzerland}
\affiliation{AGH University of Krakow, Faculty of Physics and Applied Computer Science, 30-059 Krakow, Poland}

\author{R. Frison}
\affiliation{Physik-Institut, Universit\"{a}t Z\"{u}rich, Winterthurerstrasse 
190, CH-8057 Z\"{u}rich, Switzerland}

\author{A. Morawietz}
\affiliation{Physik-Institut, Universit\"{a}t Z\"{u}rich, Winterthurerstrasse 
190, CH-8057 Z\"{u}rich, Switzerland}

\author{L. Martinelli}
\affiliation{Physik-Institut, Universit\"{a}t Z\"{u}rich, Winterthurerstrasse 
190, CH-8057 Z\"{u}rich, Switzerland}

\author{J.~Choi}
\affiliation{Diamond Light Source, Harwell Campus, Didcot OX11 0DE, United Kingdom}

\author{D. Bucher}
\affiliation{Physik-Institut, Universit\"{a}t Z\"{u}rich, Winterthurerstrasse 190, CH-8057 Z\"{u}rich, Switzerland}

\author{O. Ivashko}
\affiliation{Deutsches Elektronen-Synchrotron DESY, Notkestra{\ss}e 85, 22607 Hamburg, Germany}

\author{M.~v.~Zimmermann}
\affiliation{Deutsches Elektronen-Synchrotron DESY, Notkestra{\ss}e 85, 22607 Hamburg, Germany}

\author{N.~B.~Christensen}
\affiliation{Department of Physics, Technical University of Denmark, DK-2800 Kongens Lyngby, Denmark}

\author{D.~G.~Mazzone}
\affiliation{Laboratory for Neutron Scattering and Imaging, Paul Scherrer Institut, CH-5232, Villigen-PSI, Switzerland}

\author{G. Simutis}
\affiliation{Laboratory for Neutron and Muon Instrumentation, Paul Scherrer Institut, CH-5232 Villigen PSI, Switzerland}

\author{A. A. Turrini}
\affiliation{Laboratory for Neutron Scattering and Imaging, Paul Scherrer Institut, CH-5232, Villigen-PSI, Switzerland}

\author{L. Thomarat}
\affiliation{Laboratory for Neutron and Muon Instrumentation, Paul Scherrer Institut, CH-5232 Villigen PSI, Switzerland}
\affiliation{École normale supérieure Paris-Saclay, 91190 Gif-sur-Yvette, France}

\author{D. W. Tam}
\affiliation{Laboratory for Neutron Scattering and Imaging, Paul Scherrer Institut, CH-5232, Villigen-PSI, Switzerland}

\author{M. Janoschek}
\affiliation{Physik-Institut, Universit\"{a}t Z\"{u}rich, Winterthurerstrasse 
190, CH-8057 Z\"{u}rich, Switzerland}
\affiliation{Laboratory for Neutron and Muon Instrumentation, Paul Scherrer Institut, CH-5232 Villigen PSI, Switzerland}

\author{T.~Kurosawa}
\affiliation{Department of Physics, Hokkaido University, Sapporo 060-0810, Japan}

\author{N.~Momono}
\affiliation{Department of Physics, Hokkaido University, Sapporo 060-0810, Japan}
\affiliation{Department of Applied Sciences, Muroran Institute of Technology, Muroran 050-8585, Japan}

\author{M.~Oda}
\affiliation{Department of Physics, Hokkaido University, Sapporo 060-0810, Japan}

\author{Qisi~Wang}
\email{qwang@cuhk.edu.hk}
\affiliation{Physik-Institut, Universit\"{a}t Z\"{u}rich, Winterthurerstrasse 
190, CH-8057 Z\"{u}rich, Switzerland}
\affiliation{Department of Physics, The Chinese University of Hong Kong, Shatin, Hong Kong, China}

\author{J.~Chang}
\affiliation{Physik-Institut, Universit\"{a}t Z\"{u}rich, Winterthurerstrasse 190, CH-8057 Z\"{u}rich, Switzerland}



\maketitle

\section{Abstract} 
\textbf{
Unconventional superconductivity often couples to other electronic orders in a cooperative or competing fashion. 
Identifying external stimuli that tune between these two limits is of fundamental interest. 
Here, we show that strain perpendicular to the copper-oxide planes couples directly to the competing interaction between charge stripe order and superconductivity in La$_{1.88}$Sr$_{0.12}$CuO$_4$ ({LSCO}). Compressive $c$-axis pressure 
amplifies stripe order within the superconducting state, while having no impact on the normal state. By contrast, strain dramatically diminishes the magnetic field enhancement of stripe order in the superconducting state. These results suggest that $c$-axis strain acts as tuning parameter of the competing interaction between charge stripe order and superconductivity. This interpretation implies a uniaxial pressure-induced ground state in which the competition between charge order and superconductivity is reduced. 
}\\ [2mm]

\section{Introduction} 
Electronic phases may coexist microscopically, either in a collaborative or competing manner. In elementary chromium, for example, spin and charge density wave orders collaboratively coexist with commensurate ordering vectors~\cite{Gibbs1988,Jiang_1997,Hu2022}. A similar spin-charge intertwined order is found in doped lanthanum-based (La-based) cuprate superconductors~\cite{Tranquada1994,TranquadaPRB96,HuckerPRB2011,ThampyPRB2014,AchkarSCI2016}.
Competing interaction is often found in the 
context of unconventional superconductivity. 
For example, in kagome metals \cite{Yu2021,Song2021}, pnictides \cite{NandiPRL2010, Hu_PhysRevB.101.020507, Allred_PhysRevB.90.104513}, and heavy Fermion systems \cite{Flouquet_2011}, superconductivity can be optimized through the suppression of charge or spin density wave orders.

However, an interplay between density waves and superconductivity -- at least theoretically -- can lead to a collaborative state. This state would be  characterized by a spatially modulated Cooper pair density with a commensurate wave vector. Extensive experimental and theoretical efforts have been devoted to study this novel superconducting state \cite{Agterberg2020}. Theory works have predicted a connection between superconductivity and stripe order through a so-called pair density wave \cite{WangPRB2018}. Signatures of these pair density waves have been reported by scanning tunneling microscopy \cite{DaiPRB2018}, but direct diffraction evidence is still missing. 
A general challenge is therefore to switch the coupling between superconductivity and charge order from competing to collaborative. Ideally, an external stimulus would tune the coupling between these two phases.

Here, using high-energy x-ray diffraction, we show how compressive $c$-axis uniaxial pressure, perpendicular to the copper-oxide planes, enhances stripe order inside the superconducting state of La$_{1.88}$Sr$_{0.12}$CuO$_4$ (LSCO), while charge order remains unchanged in the normal state. We furthermore discover that the magnetic field enhancement of charge order  inside the superconducting state  is dramatically reduced upon compressive $c$-axis strain application.
This observation suggests a correspondingly reduced phase competition. 
We thus demonstrate that $c$-axis pressure acts directly on the coupling
between charge stripe order and superconductivity.

\begin{figure*}
	\begin{center}
		\includegraphics[width=0.948\textwidth]{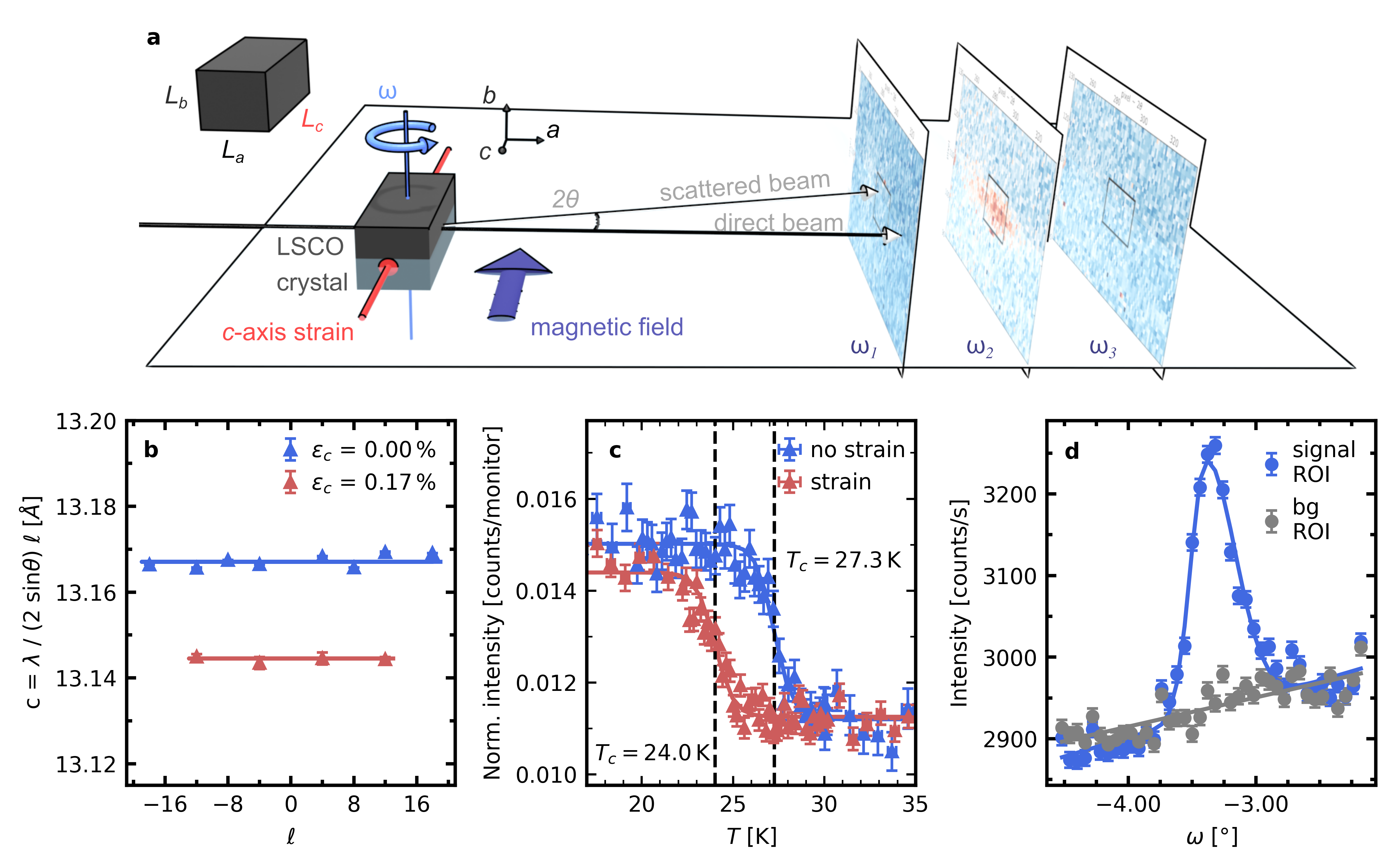}
	\end{center}
 	\caption{\textbf{Uniaxial pressure application in the hard x-ray diffraction experiment on LSCO. a} Schematic illustration of the scattering geometry for high-energy x-ray diffraction on LSCO. Uniaxial pressure and magnetic field are applied along the crystallographic $c$-axis direction on a cuboid-shaped crystal
  with dimensions $L_a \cdot L_b \cdot L_c$ as sketched in the top left inset. Detector read-outs with exemplary region-of-interest (ROI, black box in detector images) for three different rotation angles $\omega$ around the vertical axis are shown.
 	\textbf{b} $c$-axis lattice parameters, extracted from fits of (0, 0, $\ell$) Bragg peaks, corresponding to $c$-axis strains $\varepsilon_c$. Solid lines are least-square fits to $c = \lambda / (2 \sin{\omega}) \cdot \ell$, where $\lambda$ is the wavelength of the x-rays. Error bars are standard deviations obtained from the fitting procedure.
 \textbf{c} Spin flip intensities from polarized neutron scattering  at $\boldsymbol{\tau}=(-1, -1, 0)$ on LSCO ($T_c=27$\,K), normalized by the monitor and scaled to the integrated Bragg peak intensity of the unstrained measurement. The drop of the curves yields the superconducting transition temperature with and without $c$-axis uniaxial pressure (see Methods for detailed definition of $T_c$).
\textbf{d} Charge order peak at 30\,K obtained by integrating the intensities in a region-of-interest (ROI) around $\textbf{Q}_{co}=(0.231, 0, 12.5)$ (blue points).  A background (bg, grey points) is estimated
by a similar integration of a ROI slightly shifted  off the $(h,0,\ell)$ scattering plane. Error bars in c and d are dictated by counting statistics.}
\label{fig:exp}
\end{figure*}

\section{Results} 
\textbf{X-ray Methodology.}\\
Stripe charge order in La-based cuprates manifests itself by weak reflections at $\bf{Q}_{co}=\boldsymbol{\tau}$+$(\delta, 0, 0.5)$ where $\boldsymbol{\tau}$ represents fundamental Bragg peaks and $\delta\approx 1/4$ is the stripe incommensurability \cite{tranquada, HuckerPRB2011, CroftPRB2014, ThampyPRB2014}. We adopted an
 x-ray transmission geometry with crystalline $a$- and $c$-axes spanning the horizontal scattering plane as illustrated in Fig.\,\ref{fig:exp}a. Magnetic field and uniaxial pressure were applied along the $c$-axis direction. 
 
 \textbf{Uniaxial Pressure Application.} \\
 The strain as a result of uniaxial $c$-axis pressure can be directly estimated from lattice parameter measurements. Pressure-induced compression of the $c$-axis lattice parameter is evidenced by a shift of $(0, 0, n$) Bragg peaks to larger scattering angles. 
 Precise strain characterization utilizes multiple such Bragg peaks 
 with 
 $n$ being an even integer
 (see Method section and Fig.\,\ref{fig:exp}b). The resulting $c$-axis lattice parameter for strained and unstrained LSCO is exemplified in Fig.\,\ref{fig:exp}b. 
 As expected, the uniaxial $c$-axis pressure reduces the $c$-axis lattice parameter that in turn lowers the superconducting transition temperature $T_c$ \cite{PavariniPRL2001, Gugenberger_prb1994, Jalekeshow_physicac2023}.
 Using polarized neutron scattering 
 we confirmed the decrease of $T_c$ with compressive $c$-axis strain~\cite{Nakamura1999,Sato1997,LocquetActaP1997,Takeshita2004}
-- see Fig.\,\ref{fig:exp}c and Method section. Exploiting that the lower critical field for superconductivity $H_{c1}$ is low, we track the excess depolarisation of the neutron beam due to flux trapped along the $c$-axis after field-cooling through $T_C$. The in-plane polarized neutrons are depolarized by this trapped flux. Upon crossing $T_c$, the flux is released. 

 \begin{figure*}
	\begin{center}
		\includegraphics[width=\textwidth]{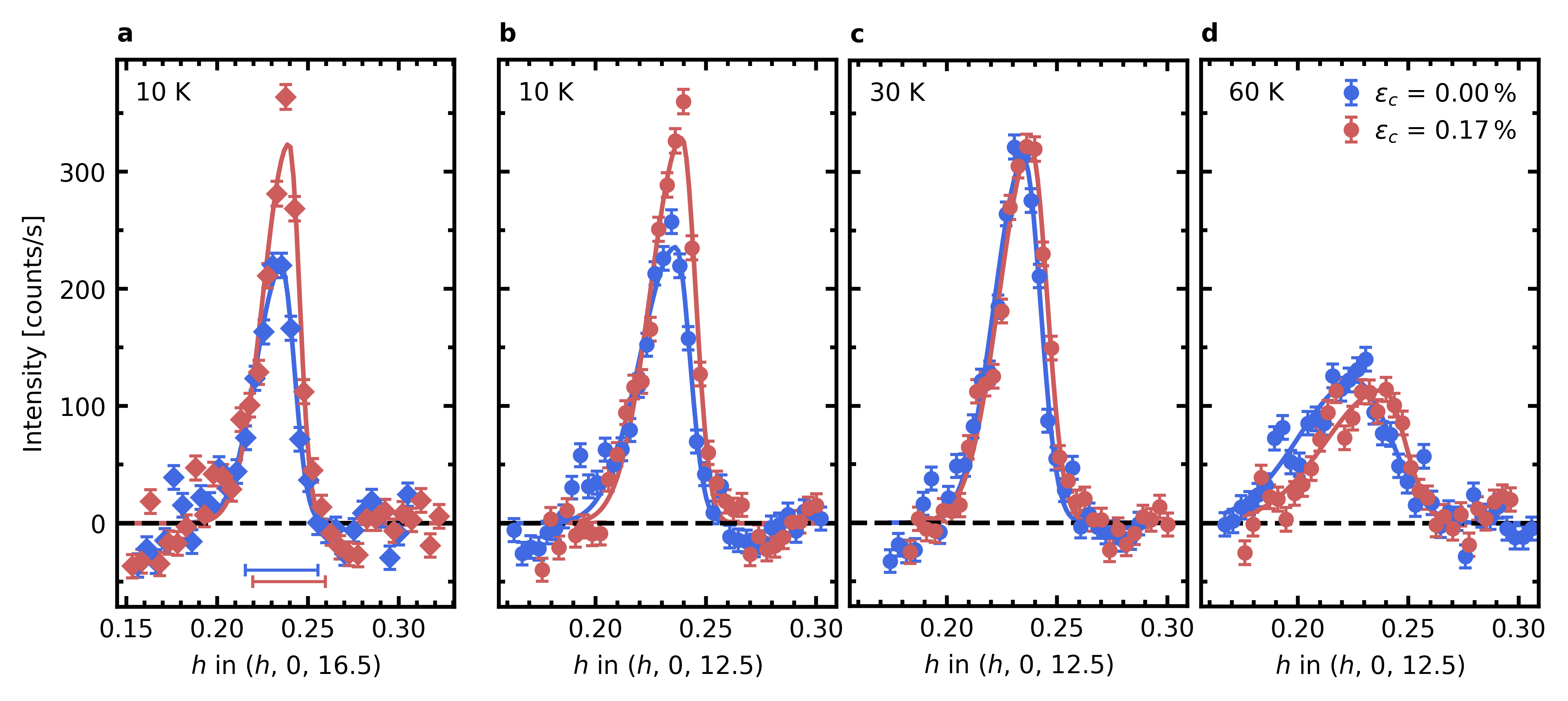}
	\end{center}
 	\caption{\textbf{Charge order reflection in LSCO upon application of $c$-axis uniaxial strain.} 
  \textbf{a-d} Background subtracted charge order reflections for temperatures and momenta as indicated. Red (blue) points are recorded with (without) compressive $c$-axis pressure application.
  Error bars stem from counting statistics and solid lines are fits with a split-normal distribution including a 
  linear background. The horizontal lines in the bottom of (a) -- applying to all panels --   represent a systematic error in $h$ stemming from twinning of the sample \cite{ChoiPRL2022}.} 
	\label{fig2:chargeordertemps}
\end{figure*}

\textbf{Uniaxial $c$-axis strain effects on charge order.}\\
X-ray diffraction intensity was collected using a two-dimensional single-photon detector. 
Detector regions-of-interest (ROI) are defined such that the signal or background of interest is covered (see Supplementary Fig.\,1). We constructed standard one-dimensional rocking curves (see Fig.\,\ref{fig:exp}d for $\bf{Q}_{co}\rm=(\delta, 0, 12.5)$).  An advantage of 2D-detectors (over point detectors) is that a background can be estimated by slightly shifting the ROI (grey data in Fig.\,\ref{fig:exp}d).

\begin{figure*}
	\begin{center}
    \includegraphics[width=0.98\textwidth]{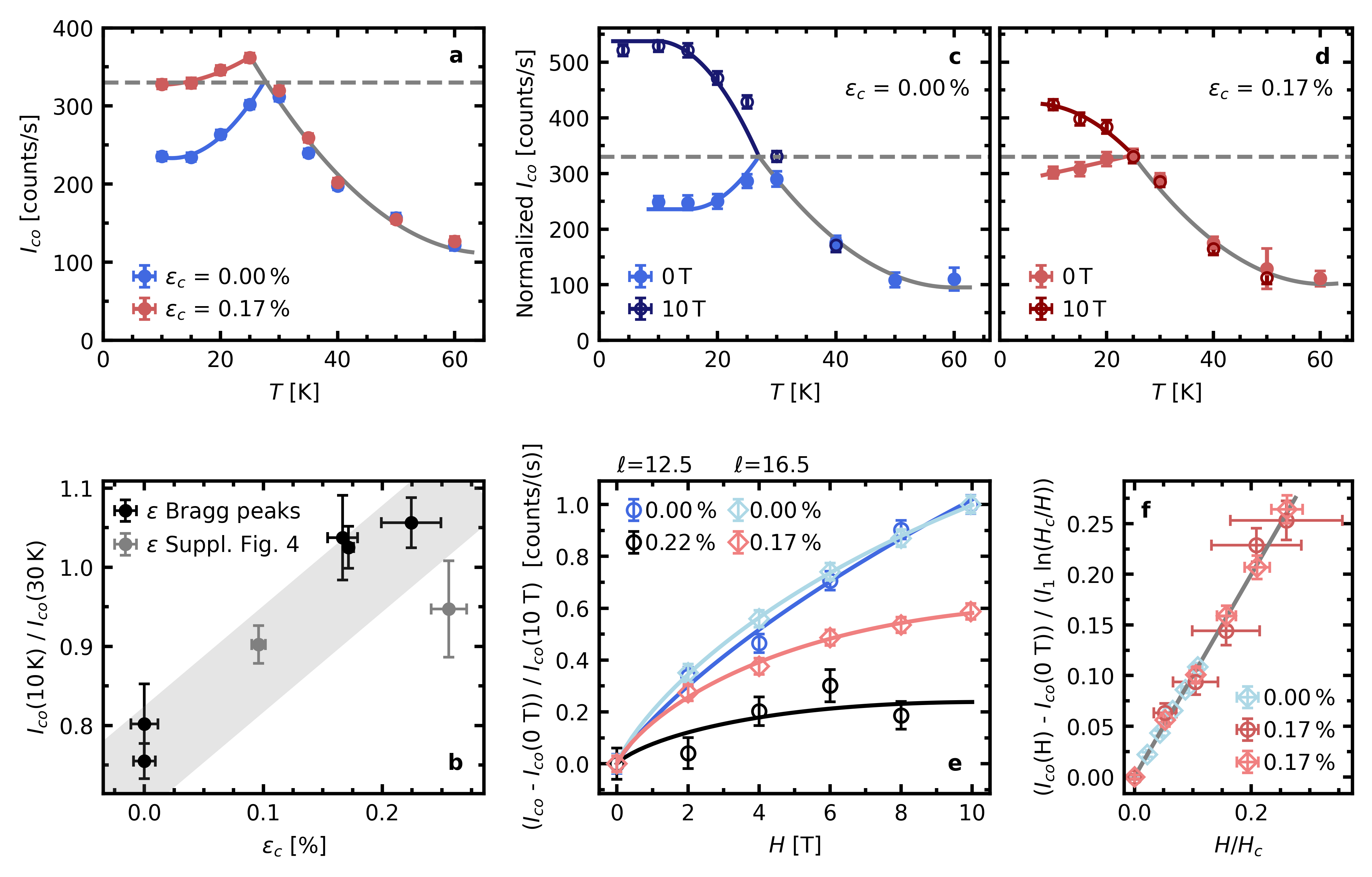}
	\end{center}
    \caption{\textbf{Phase competition of charge order and superconductivity.}
    \textbf{a} Temperature dependence of the amplitude $I_{co}$ of the (0.231, 0, 12.5) charge order peak without and with strain measured in a setup without magnetic field. 
    \textbf{b} Low-temperature (10\,K) charge order amplitude normalized to the normal state (30\,K) plotted as a function of $c$-axis strain. For the black (grey) points, 
    the strains are directly (indirectly) measured. Indirect measurements use the 
    calibration curve in Supplementary Fig.\,4. 
    The grey shaded area is a guide to the eye.
    \textbf{c, d} Temperature dependence for charge order peak at (0.231, 0, 12.5) in 0\,T and 10\,T without and with strain, respectively measured in the cryomagnet. The horizontal dashed lines at $~I_{\text{co}}(T_C)$ in panels a, c, d serve to facilitate comparison between experiments.
    \textbf{e} Magnetic field dependence at 10\,K of the charge order peak amplitude at (0.231, 0, 12.5) (circles) and (0.231, 0, 16.5) (diamonds) for strains as indicated. Solid lines are fits to the data with $I_{co}(\mathcal{H})=I_{co}(0)+I_1 \mathcal{H} \ln(1/\mathcal{H})$ \cite{DemlerPRL2001}, where $\mathcal{H}=H/H_c$, $I_1$ is a fitting parameter, and $H_c$ is a critical field scale.
    \textbf{f} Comparison of data obtained with and without strain at (0.231, 0, 12.5) (circles) and (0.231, 0, 16.5) (diamonds) ($T=10$\,K). By application of uniaxial pressure, we roughly double the reachable parameter space in $H/H_c$. 
    Solid lines in a-c are guides to the eye. The dashed lines indicate the charge order amplitude at $T_c$ and ambient pressure. Error bars are standard deviations from the respective fits.}
	\label{fig:magnfield}
\end{figure*}

In Fig.\,\ref{fig2:chargeordertemps}, we show scans through $\bf{Q}_{co}$=($\delta$, 0, 12.5) and $(\delta, 0, 16.5)$, with and without uniaxial $c$-axis pressure. Data for a La$_{2-x}$Sr$_{x}$CuO$_4$ crystal with slightly different doping are shown in Supplementary Fig.\,2. Intensities and fits are presented after subtracting the background.
In the normal state ($T>T_c$), no 
pressure effect on charge stripe order is observed. Further, we find a significant pressure-induced enhancement of the charge order reflection  inside the superconducting state.
The correlation length and incommensurability $\delta$ remain virtually unaffected by uniaxial pressure. Observed shifts are within the error bars and thus negligible. From here, we therefore consider the charge order peak amplitude as a function of temperature, uniaxial $c$-axis pressure, and magnetic field. The peak amplitude $I_{co}$ is extracted by fitting intensity profiles with a split-normal distribution on a linear background -- see Fig.\,\ref{fig:exp}d, \ref{fig2:chargeordertemps},  Supplementary Fig.\,3 and Methods.
\pagebreak
The temperature dependence of the charge order amplitude is shown in Fig.\,\ref{fig:magnfield}a for strained and unstrained conditions. In the absence of a magnetic field, compressive $c$-axis pressure enhances the charge order inside the superconducting state. The charge order peak amplitude, due to phase competition \cite{CroftPRB2014}, displays
a cusp at $T_c$. The cusp is shifted to slightly lower temperatures upon application of $c$-axis pressure. Assuming phase competition between charge order and superconductivity, this suggests a reduction of the superconducting transition temperature, in agreement with the measurements in Fig.\,\ref{fig:exp}c. At our base temperature ($T = 10$\,K), the relative peak amplitude, $I_{co}(10\,\text{K}) / I_{co}(30\,\text{K})$, scales approximately linearly with the applied strain $\varepsilon_c$ (see Fig.\,\ref{fig:magnfield}b). Within the examined range of  $\varepsilon_c$, the charge order peak amplitude increases by about 25\,\%.

\textbf{Magnetic field effect.}\\ 
Without strain, magnetic field effects on charge and spin order inside the superconducting state have already been studied~\cite{Christensen14,ChoiPRL2022,ChangPRB2008,LakeNature2002,HueckerPRB2013,Khaykovich_PRB2005}. Consistent with previous studies, we find an increase of the charge order amplitude for $T<T_c$ (see Fig.\,\ref{fig:magnfield}c). 
At base temperature, the charge order intensity scales approximately linearly with magnetic field. At 10\,T, the peak amplitude is more than doubled. This is in strong contrast to the ground state reached through application of $c$-axis pressure. Here, as depicted in Fig.\,\ref{fig:magnfield}d, 
a much weaker field effect is observed. The magnetic field effect for strained and unstrained conditions is shown in Fig.\,\ref{fig:magnfield}e. \pagebreak Upon increasing 
strain $\varepsilon_c$, 
a decreasing magnetic field effect on the charge order intensity is observed inside the superconducting state.



\section{Discussion} 
\begin{figure}
	\begin{center}
    \includegraphics[width=0.49\textwidth]{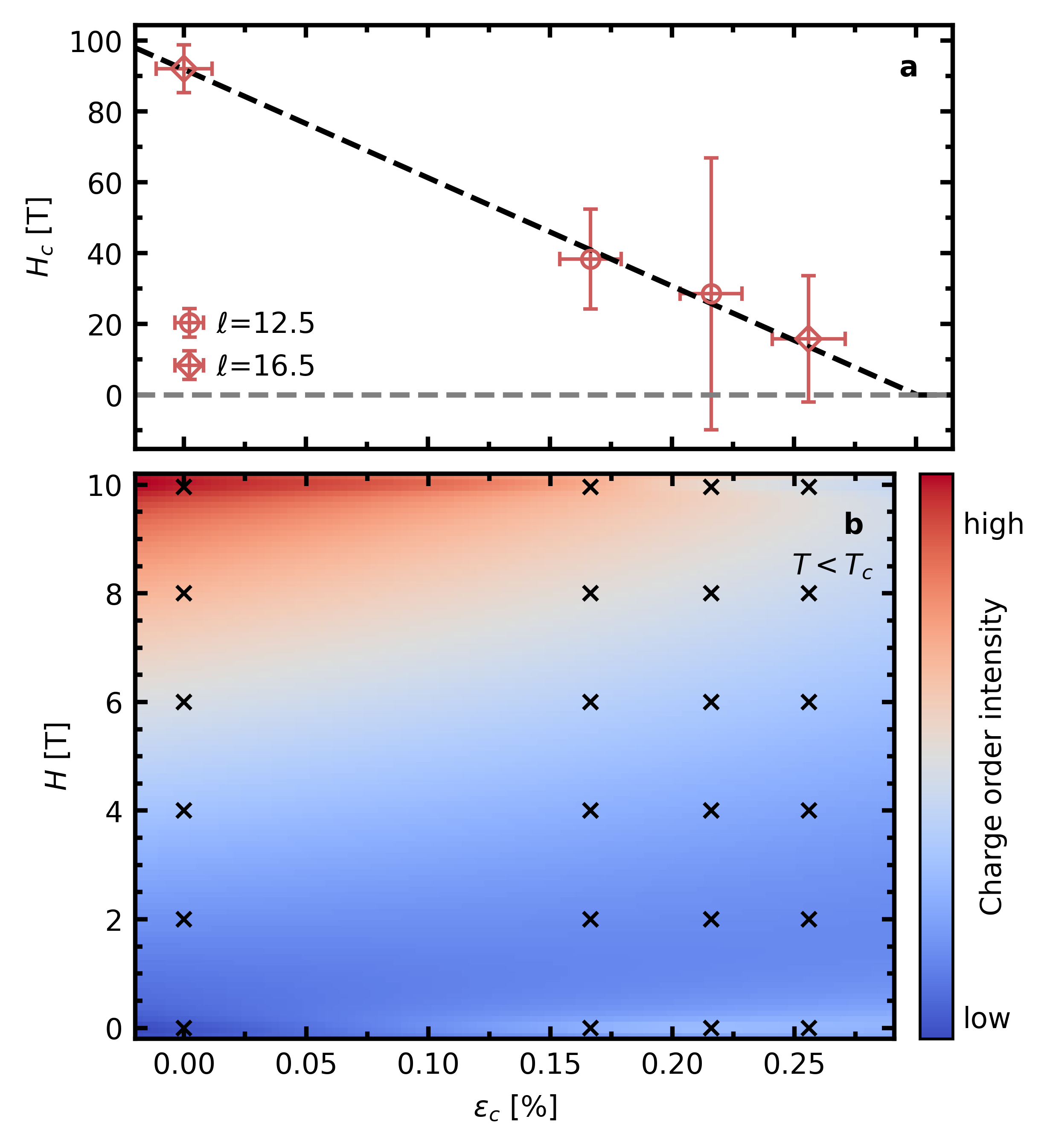}
	\end{center}
    \caption{\textbf{Magnetic field and strain effects.} \textbf{a} The magnetic field scale $H_c$ versus compressive $c$-axis strain $\varepsilon_c$ at 10\,K. Data points were obtained from measurements and fits of charge order peaks at $(0.231, 0, \ell)$ with $\ell = 12.5, 16.5$. Vertical error bars are Gaussian standard deviations from fits to the magnetic field dependence of the charge order reflection. Horizontal error bars reflect uncertainty of the strain estimation. The black dashed line is a linear guide to the eye 
     \textbf{b} Schematic illustration of the magnetic field $H$ and strain effect on the charge order peak amplitude inside the superconducting state. 
     We expect -- by extrapolation -- the magnetic field enhancement to fully vanish. Crosses display the phase space covered by our experiments.}
	\label{fig:phasediag}
\end{figure}

There has been an interest to study the stripe order beyond the upper critical field ($H_{c2}$) for superconductivity~\cite{WenNatComm2023}.
The magnetic field dependence of the charge order amplitude in Fig.\,\ref{fig:magnfield}e is often modelled by $I_{co}(\mathcal{H})=I_{co}(0)+I_1 \mathcal{H} \ln(1/\mathcal{H})$, where $\mathcal{H}=H/H_c$ and $I_1$ is a fitting parameter. The critical field scale $H_c$ has been linked to the upper critical field $H_{c2}$ of superconductivity~\cite{DemlerPRL2001}. We however stress that these two field scales ($H_{c}$ and $H_{c2}$) are not necessarily identical.
We 
show in Fig.\,\ref{fig:magnfield}f 
that we could double the reachable parameter space in $\mathcal{H}$ by applying strain along $c$.
By fitting the observed charge order peak amplitude versus magnetic field, we extract $H_c$ at base temperature (see fitted curves in Fig.\,\ref{fig:magnfield}e). In Fig.\,\ref{fig:phasediag}a, we plot $H_c$ versus strain $\varepsilon_c$.
The field scale $H_c$ seems to scale approximately linearly with $c$-axis strain: $H_c=H_c(0)-\alpha \varepsilon_c$ with $\alpha$ being a constant. By extrapolation, our results suggest that there exists a critical $c$-axis strain at which $H_c\rightarrow 0$. At this critical strain $\varepsilon_{crit}=H_c(0)/\alpha$ $\approx 0.3$\,\%, 
the magnetic field effect is expected to vanish.

Outlining our findings, we illustrate schematically in Fig.\,\ref{fig:phasediag}b the charge order peak amplitude versus magnetic field and $c$-axis strain inside the superconducting state.
The region with highest intensities is found at high magnetic fields and low strains. The strength of the magnetic field effect depends on the uniaxial strain along $c$. At the highest applied $c$-axis strain, the magnetic field effect is strongly reduced. Typically, magnetic field effects inside the superconducting state are interpreted in terms of \textit{(i)} vortex physics \cite{LakeScience2001, WuNatComm2013} or \textit{(ii)} phase competition. 
The vortex density and volume increase with increasing magnetic field strength. 
It is however difficult to explain the strain-field effects in terms of vortices alone.
Within the framework of vortex physics~\cite{LakeScience2001, WuNatComm2013}, the magnetic field induces normal state vortices. The field-enhanced charge order may then stem from real-space volume increase. However, we see that in zero-field, $c$-axis strain enhances the charge order peak amplitude. This strongly suggests that vortex physics alone cannot account for the observed phenomenology. 
Even in the high-magnetic field state where vortices are present, it is not obvious how to explain the strain effect without also involving phase competition. 

Reduced (absence of) magnetic field effect would imply either that one phase is partially suppressed (absent) or that the competing interaction is weakened. 
As no pressure effect is found in the normal state, charge order itself is inert to $c$-axis pressure. 
On the other hand, superconductivity is known to be weakened by increasing compressive $c$-axis pressure~\cite{NakamuraPRB2000}.
One 
commonly accepted explanation is that compressive $c$-axis pressure reduces the apical oxygen distance to the CuO$_2$ layers. This change of crystal field environment, in turn, boosts the hybridization of \dx\ and \dz\ orbitals \cite{MattNatCommun2018,KramerPRB2019}. Theoretically, such hybridization is expected to be unfavorable for $d$-wave superconductivity~\cite{HirofumiPRL10,HirofumiPRB12}. Therefore, $c$-axis pressure diminishes the superconducting order parameter. However, 
there is no evidence of superconductivity being completely suppressed upon moderate pressure along the $c$-axis~\cite{PavariniPRL2001,Nakamura1999}. 
Superconductivity and charge order are therefore expected to coexist. Our results thus suggests a pressure-induced ground state in which the phase competition between charge order and superconductivity is reduced. 

\section{Methods} 
\subsection{Crystal Growth and Characterization}
Our La$_{2-x}$Sr$_x$CuO$_4$\; single crystals were synthesised using the traveling solvent floating zone method.
The growths were carried out in a $2.0$\,atm oxygen pressure. Post-growth annealing was carried out in an 800\,degree oxygen atmosphere for 100\,hours. For a very similar growth recipe, superconducting volume fractions were systematically studied using magnetization and specific heat~\cite{Koike2008}.
In our study, we used single crystals from three different growth cycles -- all with nominal strontium content of $x\approx 0.12$. The effective strontium contents were gauged from magnetic susceptibility measurements of the superconducting transition temperature. All x-ray diffraction data in the main text stem from 
batch I -- see Table\,\ref{tab:tab1}. For the $a$-axis, pointing approximately along the (transmitted) x-ray beam direction, the ideal sample thickness $L_a$ matches the half-value thickness ($d_{1/2}\sim$ 0.63\,mm for LSCO~\cite{absorptioncoeff, McMaster_1970CompilationOX}). At this trade-off point ($L_a\approx d_{1/2}$), half of the incident x-ray flux is transmitted through the sample. We optimized $L_a$ for the XRD experiments, so that $L_a \approx d_{1/2}$ (see Fig.\,\ref{fig:exp}a and Table\,1).

\subsection{X-Ray Diffraction and Pressure Cell}
Hard x-ray diffraction experiments were carried out at the P21.1 beamline at PETRA III (DESY, Hamburg) synchrotron using 101.6\,keV photons. 
All measurements were performed with a Dectris Pilatus 100K CdTe detector. Uniaxial pressure and magnetic field were applied along the crystallographic $c$-axis. 
Misalignment of $c$-axis and press direction is no more than $4^\circ$. As such, unintentional in-plane pressure application is negligible \cite{ChoiPRL2022}. In the absence of a magnetic field, a standard (ex-situ) screw cell was used \cite{ChoiPRL2022,SimutisCommPhys2022}. An adaption was made to fit into a horizontal magnet~\cite{Christensen14,HueckerPRB2013} with pressure application parallel to the direction of the magnetic field. The designs of the two uniaxial pressure cells are shown in Supplementary Fig.\,5 a, b.
For both cells, the horizontal scattering plane is spanned by a copper-oxygen bond direction and the $c$-axis. This gives access to the $(h, 0, \ell)$ reciprocal plane. Consistent results are  obtained with both setups (displex and cryomagnet) -- see Supplementary Fig.\,6.
Throughout the manuscript,  reciprocal space vectors $(h,k,\ell)$ are given in units of $(\frac{2\pi}{a}, \frac{2\pi}{b}, \frac{2\pi}{c})$ with $a = b \approx 3.77$\,\AA~ and $c \approx 13.2$\,\AA. 
Notice that even though LSCO has a low-temperature orthorhombic crystal structure with $a$ and $b$ axes at 45 degrees to the Cu-O bonds with twin domains \cite{Jacobsen_PRB2015, Horibe_PRB2000, Frison_PRB2022, ChoiPRL2022}, the stripe ordering vector is traditionally expressed in tetragonal notation.
The magnitude of applied strain is determined from measurements of Bragg reflections, similar to what is described by Choi et al. \cite{ChoiPRL2022}. Bragg peak intensities are only marginally affected by the application of pressure (see Supplementary Fig.\,7 and Supplementary Tab.\,1).
We defined the zero-strain, low-temperature lattice parameter as $c_0$ and $c$-axis strain as $\varepsilon_c=(c_0-c)/c_0$. Both $c_0$ and $c$ are determined from Bragg peaks of the type $(0, 0, \pm n)$ with $n$ being an even integer (see Fig.\,\ref{fig:exp}b). Using this methodology a calibration curve between screw turn and strain is established (see Supplementary Fig.\,4).
Compressive $c$-axis strain induces -- with Poisson's ratio of $\approx 0.3$ \cite{NakamuraPRB2000} -- tensile strains along the $a$- and $b$-directions. Notice that this is different from compressive $b$-axis strain~\cite{ChoiPRL2022} that generates tensile strain along the $a$- and $c$-axes directions, again with Poisson's ratio.

\subsection{Neutron Diffraction}
Neutron diffraction data were measured at the TASP spectrometer at SINQ, PSI on LSCO $x = 0.115$ ($T_c \approx 27$\,K). The sample in the (ex-situ) pressure cell was mounted in a coil (see Supplementary Fig.\,5c), custom-built to fit inside a cryostat, which itself was mounted inside the Mu-metal polarization analysis device MuPAD \cite{Janoschek2007125}. The sample was cooled from 40\,K to 10\,K in a magnetic field of 30\,G, applied along the crystallographic $c$-direction, perpendicular to the scattering plane. At base temperature, the field was subsequently turned off. Using MuPAD’s spherical polarimetry capabilities, a full polarization matrix was measured on the in-plane Bragg peak $\boldsymbol{\tau}=(-1, -1, 0)$. 
The highest flipping ratio of 6.87 was found for neutron polarization along $y$ (in the scattering plane but perpendicular to $\boldsymbol{\tau}$). The temperature dependence of the intensity of $\boldsymbol{\tau}=(-1, -1, 0)$ was tracked in the Py-y spin flip channel upon slowly heating.
The flux trapped along the $c$-direction during the field-cooling procedure gives rise to an excess depolarization of the neutron beam, which is reflected in an increased spin-flip intensity. Upon crossing $T_c$, the trapped flux is released. 
Intensities are normalized to the monitor and scaled by the integrated Bragg peak of ($-1$, $-$1, 0) in unstrained LSCO. To extract the transition temperature,  spin flip intensities were fitted with $p_0 \cdot \text{tanh}(T - T_c) + p_1$. The resulting $T_c$ of the zero-pressure LSCO crystal is in accord with the transition temperature obtained from a magnetic susceptibility measurement of the rod (see Supplementary Fig.\,8).\\[2mm]  

\textit{Data availability.} Data are available upon request.

\textit{Code availability.} Code is available upon request.

\textit{Competing interests.} The authors declare no competing interests.

\textit{Acknowledgements:} J.K., R.F., L.M., D.B., and J.C. acknowledge support from the Swiss National Science Foundation (Projects No. 200021\_188564 and No. 200021\_296 185037). J.K. is further supported by the PhD fellowship from the German Academic Scholarship Foundation. I.B. acknowledges support from the Swiss Government Excellence Scholarship. Q.W. is supported by the Research Grants Council of Hong Kong (ECS No. 24306223), and the CUHK Direct Grant (No. 4053613). N.B.C was supported by the Danish National Council for Research infrastructure (NUFI) through DANSCATT and the
ESS-Lighthouse Q-MAT. This research was carried out at beamline P21.1 at DESY, a member of the Helmholtz Association (HGF). We would like to thank Philipp Glaevecke for technical assistance during the experiment. The research leading to this result has been supported by the project CALIPSOplus under the Grant Agreement 730872 from the EU Framework Programme for Research and Innovation HORIZON 2020. Part of the research have been carried out at the TASP endstation of the spallation source SINQ (PSI).\\[2mm]

\textit{Authors Contribution:} TK, NM, MO grew and JK prepared the LSCO crystals. JChoi, DB, MJ designed and tested the pressure cells. X-ray diffraction experiments were carried out by JK, RF, AM, OI, MvZ, NBC, QW, JChang. Polarized neutron scattering experiments were conducted by JK, NBC, DGM, GS, AAT, LT, DWT. JK carried out the data analysis with input from IB, QW, and LM. The project was conceived by JK and JChang who also wrote the manuscript with input from all authors.\\[2mm]

\begin{table}[ht!]
\vspace{3mm}
\begin{ruledtabular}
\begin{tabular}{cccc}
Crystal Batch & $T_c$ [K]& $L_a\cdot L_b\cdot L_c$ [mm$^3$] & Figure   \\ \hline
I & 27 & $0.5 \cdot 1.4 \cdot 1.4$ & \ref{fig:exp}b,d \ref{fig2:chargeordertemps}, \ref{fig:magnfield}, \ref{fig:phasediag}\\
 &  &  & Supplementary Fig. \\ 
 &  &   & 1, 3, 4, 6, 7\\
II & 27 & $0.8 \cdot 1.0 \cdot 1.5$ & \ref{fig:exp}c \\
 &  &  & Supplementary Fig.\\
 &  &   & 8 \\
III & 30 & $0.7 \cdot 0.7 \cdot 1.2$ & Supplementary Fig.\\ 
    &   & & 2\\ 
\end{tabular}
\end{ruledtabular}
\caption{\textbf{Sample dimensions.} Samples from three different batches were cut along the two Cu-O directions and the $c$-axis -- defining three lengths: $L_a$, $L_b$, and $L_c$ (see Fig.\,\ref{fig:exp}a). The last column indicates the figures which contain data taken on a given sample.}
\label{tab:tab1}	
\end{table}

\section{References}
\bibliography{Reference}

\end{document}



\title{Supplementary Material for Engineering Phase Competition Between Stripe Order and Superconductivity in \texorpdfstring{La$_{1.88}$Sr$_{0.12}$CuO$_4$}{LSCO}}

\author{J. Küspert}
\email{julia.kuespert@physik.uzh.ch}
\affiliation{Physik-Institut, Universit\"{a}t Z\"{u}rich, Winterthurerstrasse 
190, CH-8057 Z\"{u}rich, Switzerland}

\author{I. Bia\l{}o}
\affiliation{Physik-Institut, Universit\"{a}t Z\"{u}rich, Winterthurerstrasse 
190, CH-8057 Z\"{u}rich, Switzerland}
\affiliation{AGH University of Krakow, Faculty of Physics and Applied Computer Science, 30-059 Krakow, Poland}

\author{R. Frison}
\affiliation{Physik-Institut, Universit\"{a}t Z\"{u}rich, Winterthurerstrasse 
190, CH-8057 Z\"{u}rich, Switzerland}

\author{A. Morawietz}
\affiliation{Physik-Institut, Universit\"{a}t Z\"{u}rich, Winterthurerstrasse 
190, CH-8057 Z\"{u}rich, Switzerland}

\author{L. Martinelli}
\affiliation{Physik-Institut, Universit\"{a}t Z\"{u}rich, Winterthurerstrasse 
190, CH-8057 Z\"{u}rich, Switzerland}

\author{J.~Choi}
\affiliation{Diamond Light Source, Harwell Campus, Didcot OX11 0DE, United Kingdom}

\author{D. Bucher}
\affiliation{Physik-Institut, Universit\"{a}t Z\"{u}rich, Winterthurerstrasse 190, CH-8057 Z\"{u}rich, Switzerland}

\author{O. Ivashko}
\affiliation{Deutsches Elektronen-Synchrotron DESY, Notkestra{\ss}e 85, 22607 Hamburg, Germany}

\author{M.~v.~Zimmermann}
\affiliation{Deutsches Elektronen-Synchrotron DESY, Notkestra{\ss}e 85, 22607 Hamburg, Germany}

\author{N.~B.~Christensen}
\affiliation{Department of Physics, Technical University of Denmark, DK-2800 Kongens Lyngby, Denmark}

\author{D.~G.~Mazzone}
\affiliation{Laboratory for Neutron Scattering and Imaging, Paul Scherrer Institut, CH-5232, Villigen-PSI, Switzerland}

\author{G. Simutis}
\affiliation{Laboratory for Neutron and Muon Instrumentation, Paul Scherrer Institut, CH-5232 Villigen PSI, Switzerland}

\author{A. A. Turrini}
\affiliation{Laboratory for Neutron Scattering and Imaging, Paul Scherrer Institut, CH-5232, Villigen-PSI, Switzerland}

\author{L. Thomarat}
\affiliation{Laboratory for Neutron and Muon Instrumentation, Paul Scherrer Institut, CH-5232 Villigen PSI, Switzerland}
\affiliation{École normale supérieure Paris-Saclay, 91190 Gif-sur-Yvette, France}

\author{D. W. Tam}
\affiliation{Laboratory for Neutron Scattering and Imaging, Paul Scherrer Institut, CH-5232, Villigen-PSI, Switzerland}

\author{M. Janoschek}
\affiliation{Physik-Institut, Universit\"{a}t Z\"{u}rich, Winterthurerstrasse 
190, CH-8057 Z\"{u}rich, Switzerland}
\affiliation{Laboratory for Neutron and Muon Instrumentation, Paul Scherrer Institut, CH-5232 Villigen PSI, Switzerland}

\author{T.~Kurosawa}
\affiliation{Department of Physics, Hokkaido University, Sapporo 060-0810, Japan}

\author{N.~Momono}
\affiliation{Department of Physics, Hokkaido University, Sapporo 060-0810, Japan}
\affiliation{Department of Applied Sciences, Muroran Institute of Technology, Muroran 050-8585, Japan}

\author{M.~Oda}
\affiliation{Department of Physics, Hokkaido University, Sapporo 060-0810, Japan}

\author{Qisi~Wang}
\email{qwang@cuhk.edu.hk}
\affiliation{Physik-Institut, Universit\"{a}t Z\"{u}rich, Winterthurerstrasse 
190, CH-8057 Z\"{u}rich, Switzerland}
\affiliation{Department of Physics, The Chinese University of Hong Kong, Shatin, Hong Kong, China}

\author{J.~Chang}
\affiliation{Physik-Institut, Universit\"{a}t Z\"{u}rich, Winterthurerstrasse 190, CH-8057 Z\"{u}rich, Switzerland}


\maketitle

\onecolumngrid

\newcommand{\beginsupplement}{
        \setcounter{table}{0}
        \renewcommand{\thetable}{\arabic{table}}
        \renewcommand{\tablename}{\textbf{Supplementary Table}}
        \setcounter{figure}{0}
        \renewcommand{\figurename}{\textbf{Supplementary Figure}}}

\beginsupplement
\onecolumngrid

\begin{figure}[h]
\center{\includegraphics[width=\textwidth]{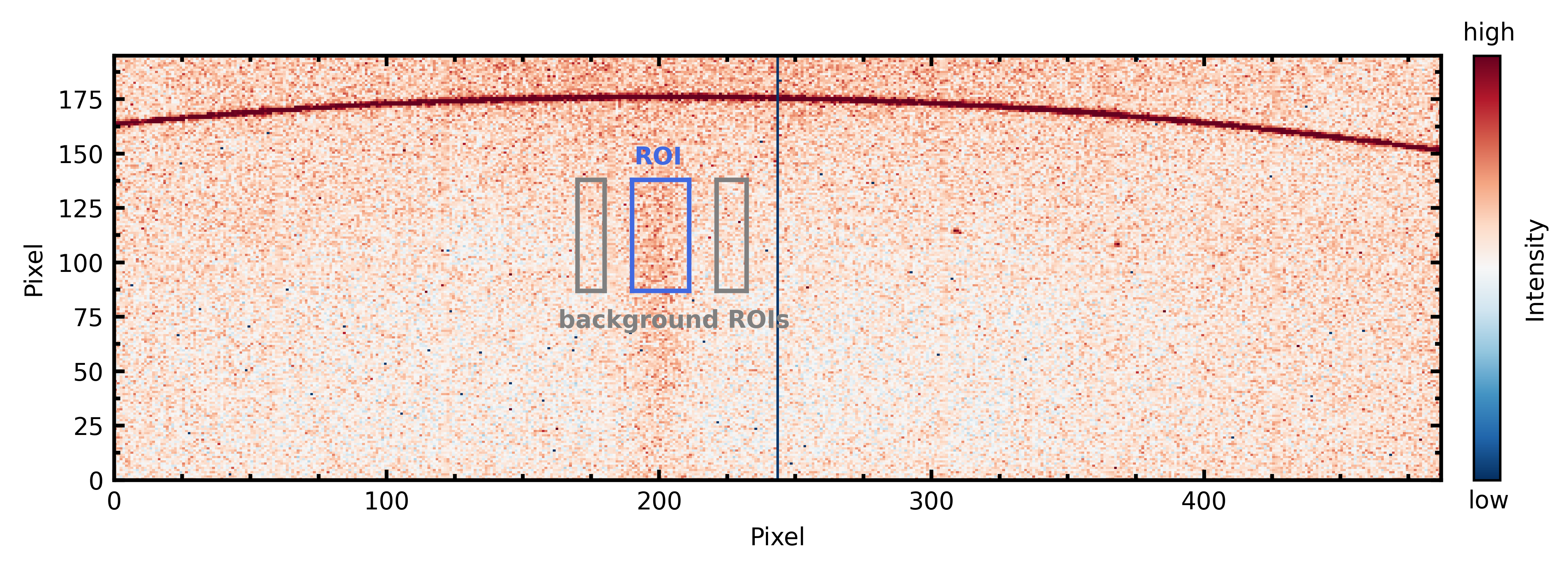}} 	
\caption{\textbf{Example of detector image.} 2D detector image recorded on LSCO $x=12$\,\% ($T_c=27$\,K) and displayed with false color scale.
Signal region-of-interest (ROI) around $\bf{Q}_{co}\rm=(0.231,0,12.5)$ and background ROIs are indicated by respectively blue and grey rectangles. The most intense feature running horizontally across the detector is a powder line stemming from the sample environment. The vertical line is a gap between two parts of the detector.}
\label{figs:frame_roi}
\end{figure} 

\begin{figure}[h]
\center{\includegraphics[width=\textwidth]{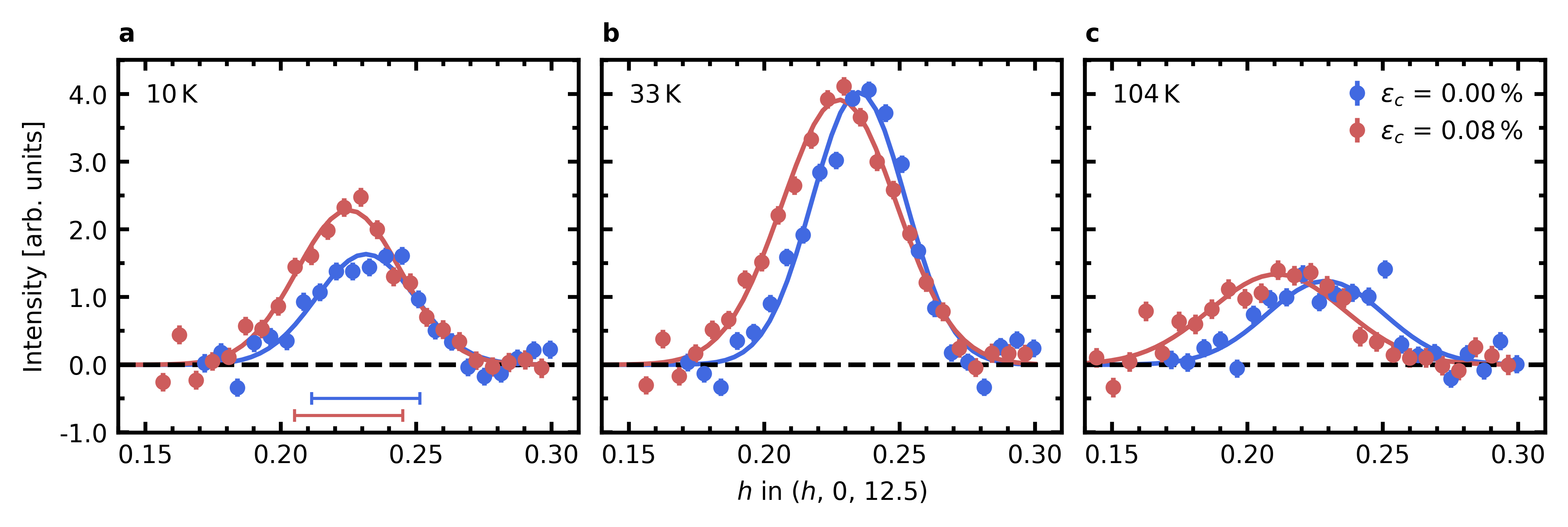}} 	
\caption{\textbf{Charge order reflection recorded on LSCO ($\bf{T_c = 30}$\,K, $\bf{x=12.5}$\,\%) upon application of $c$-axis uniaxial strain.} \textbf{a-c} Background subtracted intensity along  $\bf{Q}_{co}\rm$ $= (h, 0, 12.5)$ for temperature and strain conditions as indicated. Vertical error bars represent 
counting statistics. 
Horizontal lines in (a) -- applying to all panels --  account for a systematic error in momentum determination.
}
\label{figs:chargeorder}
\end{figure} 

\begin{figure}[h]
\center{\includegraphics[width=.5\textwidth]{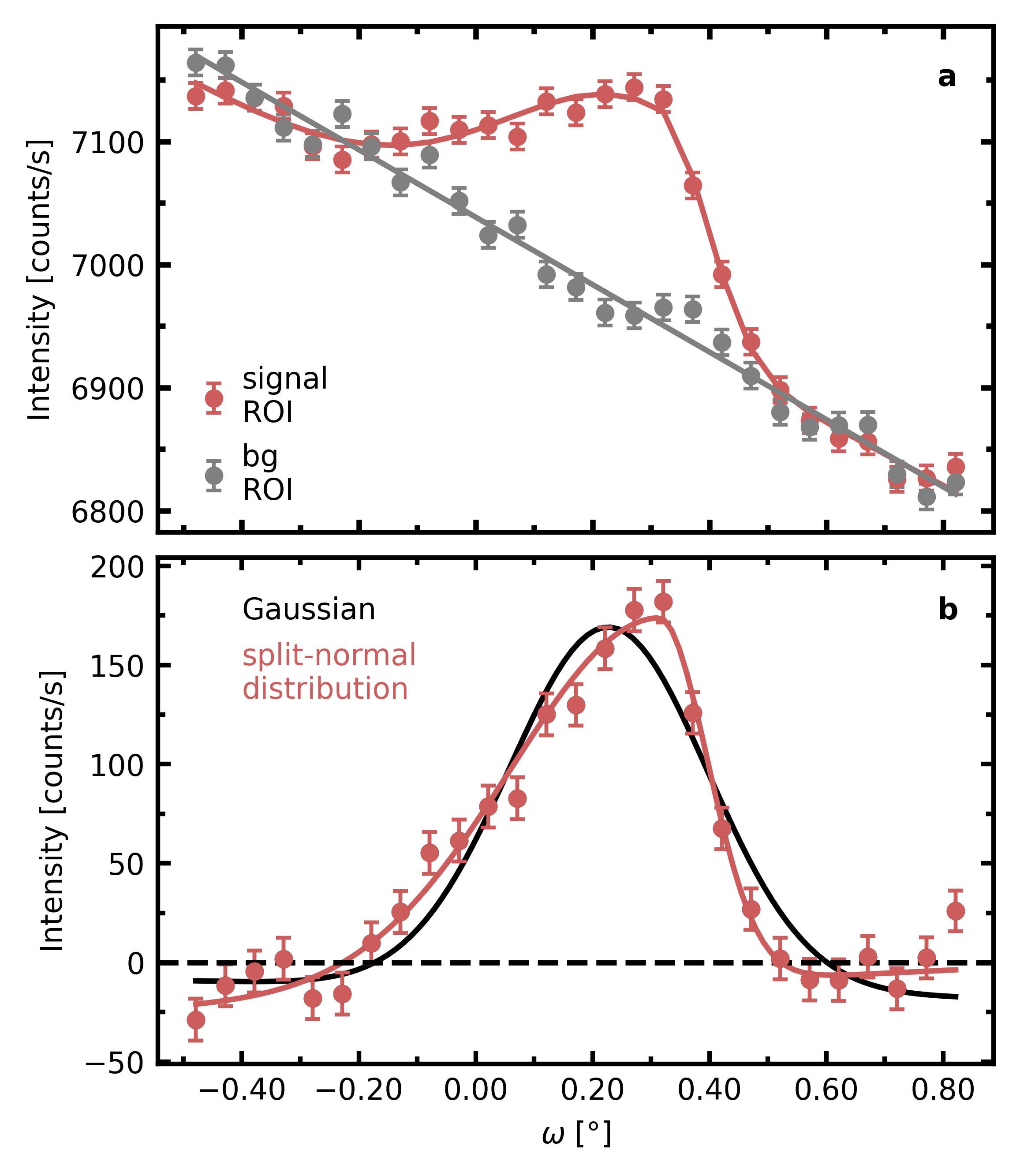}} 
\caption{\textbf{Comparison of fit functions for the charge order peak at $\bf{Q_{co}}\bf{=(0.231, 0, 12.5)}$. a} Intensity from rocking scans through the charge order reflection and corresponding background obtained from integration of "signal" and "background" ROIs (see Supplementary Fig. 1). 
The sample was exposed to an uniaxial strain of $\varepsilon_c = (0.17 \pm 0.01)$\,\%, measured by the experimental setup using a cryomagnet at 30\,K. 
 \textbf{b} Background subtracted intensities for the rocking scan shown in a. Solid black and red lines are 
respectively fits to a Gaussian function  and a split-normal distribution.
 The resulting amplitudes are $(182.99 \pm 6.00)$ counts/s and $(184.94 \pm 5.95)$ counts/s, respectively. Thus within error bars, the amplitude is independent of the two fit functions.
 As the split-normal distribution provides a better fit, this function is consistently used in our analysis.
 To be specific, a split-normal distribution with linear background is defined as: $I(\omega) = I_{co} \cdot \exp(-(\omega - \omega_0)^2 / (2 \sigma^2)) + b_1 \omega + b_0$, where $\sigma = \sigma_1$ if $\omega \leq \omega_0$ and $\sigma = \sigma_2$ if $\omega > \omega_0$. $\sigma$ is the width of the function, $\omega_0$ its peak position, $I_{co}$ the amplitude, and $b_1$ and $b_0$ fitting parameters defining the background. Error bars are set by counting statistics.}
\label{figs:comparecofits}
\end{figure} 

\begin{figure}[h]
\center{\includegraphics[width=0.5\textwidth]{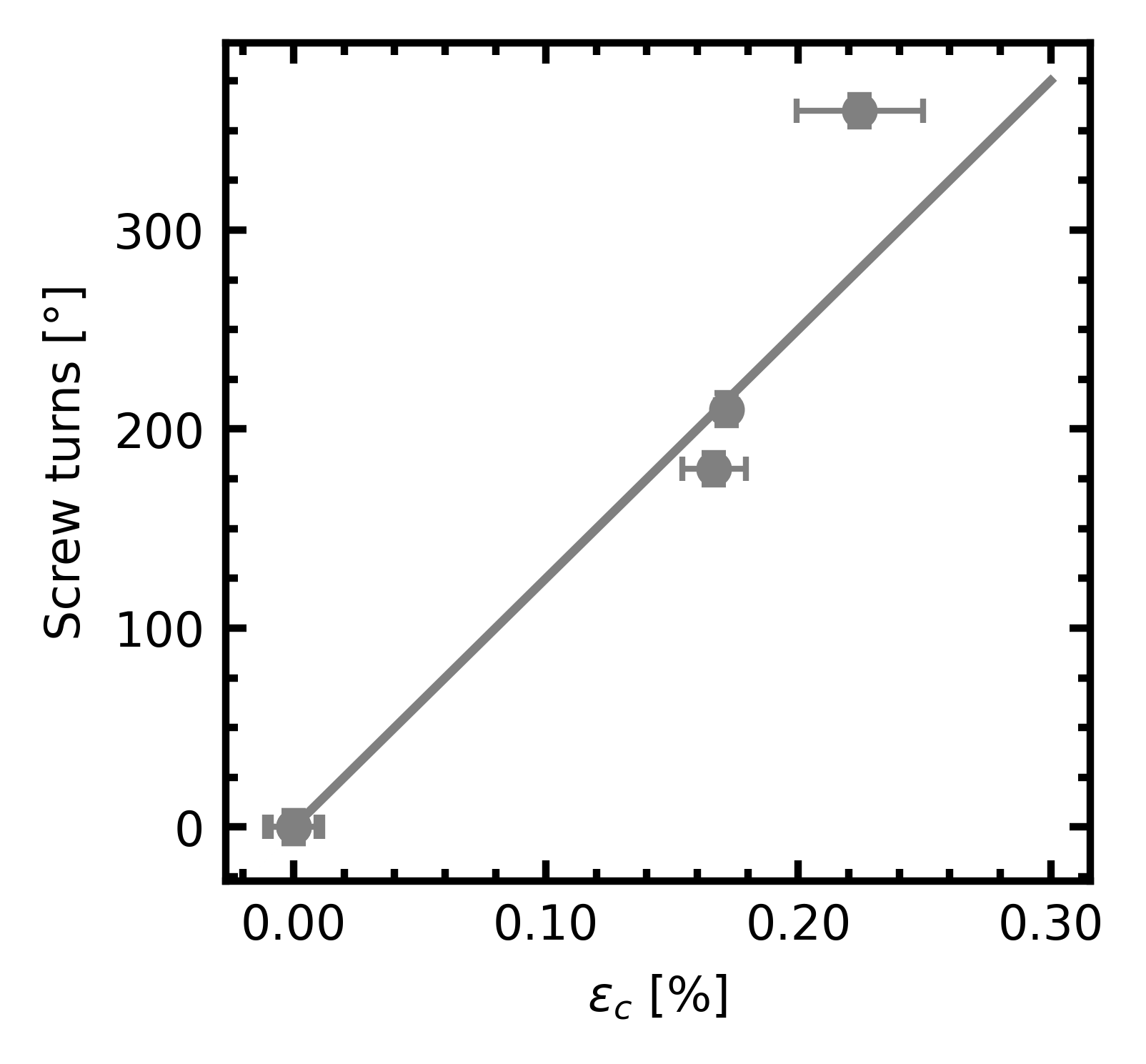}} 	
\caption{\textbf{Strain calibration curve.} Pressure cell screw turns ($st$) 
versus strain $\varepsilon_c = (c - c_0) / c_0$, where $c_0$ is the unstrained $c$-axis lattice parameter and $c$ is the strained lattice parameter. $c_0$ and $c$ are determined as described in the Methods section. The solid line is a linear fit of the form $st = m \cdot \varepsilon_c$, with $m$ being a fit parameter. Vertical errors reflect precision in screw turns ($\pm 8\,^\circ$), while horizontal errors are derived from standard deviations obtained from the fitting.}
\label{figs:strain}
\end{figure}

\begin{figure}[h]
\center{\includegraphics[width=0.75\textwidth]{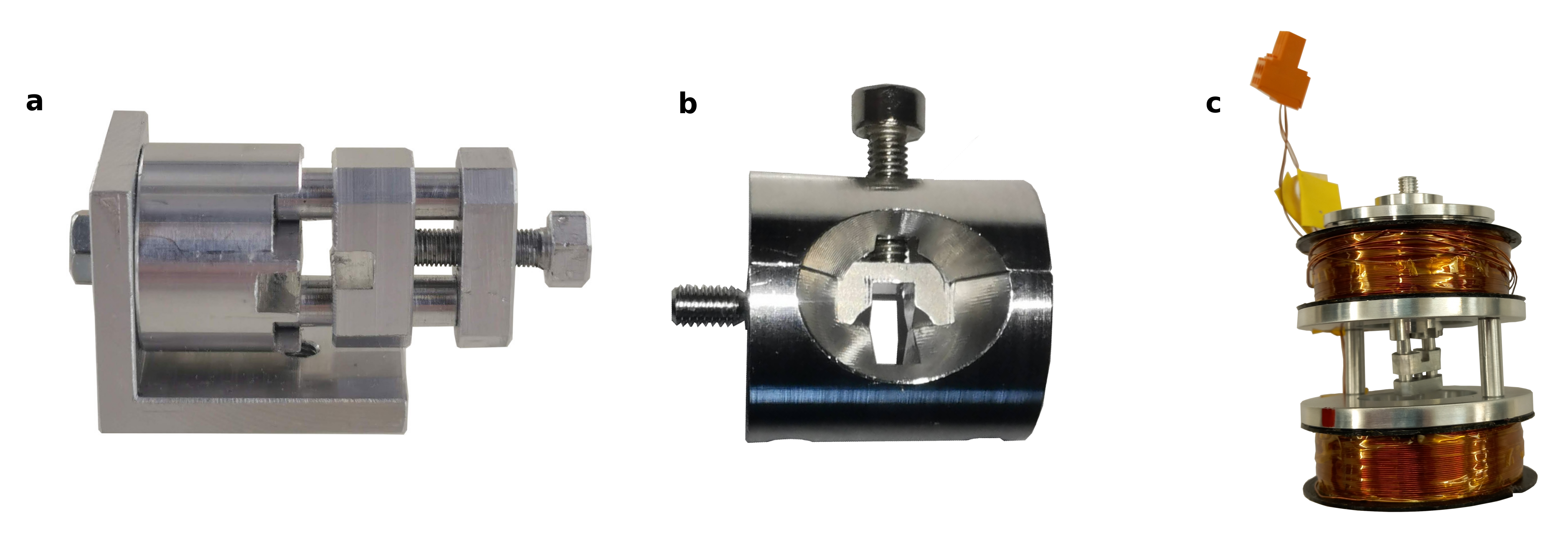}} 	
\caption{\textbf{Strain cells for $c$-axis strain application. a} Strain cell for experimental setup without magnetic field (displex cryo-environment).We extended the strain cell used in \cite{ChoiPRL2022} with an adapter to apply strain in $c$-axis direction while having $a$- and $c$-axis in the horizontal scattering plane.
\textbf{b} Strain cell for experiment with a cryomagnet. The design is  more compact in order to fit into the magnet bore.
\textbf{c} Strain cell from a (without adapter) inside the coil to be mounted inside the MuPad~\cite{Janoschek2007125} for the neutron experiment.}
\label{figs:straincells}
\end{figure}

\begin{figure}[h]
\center{\includegraphics[width=0.5\textwidth]{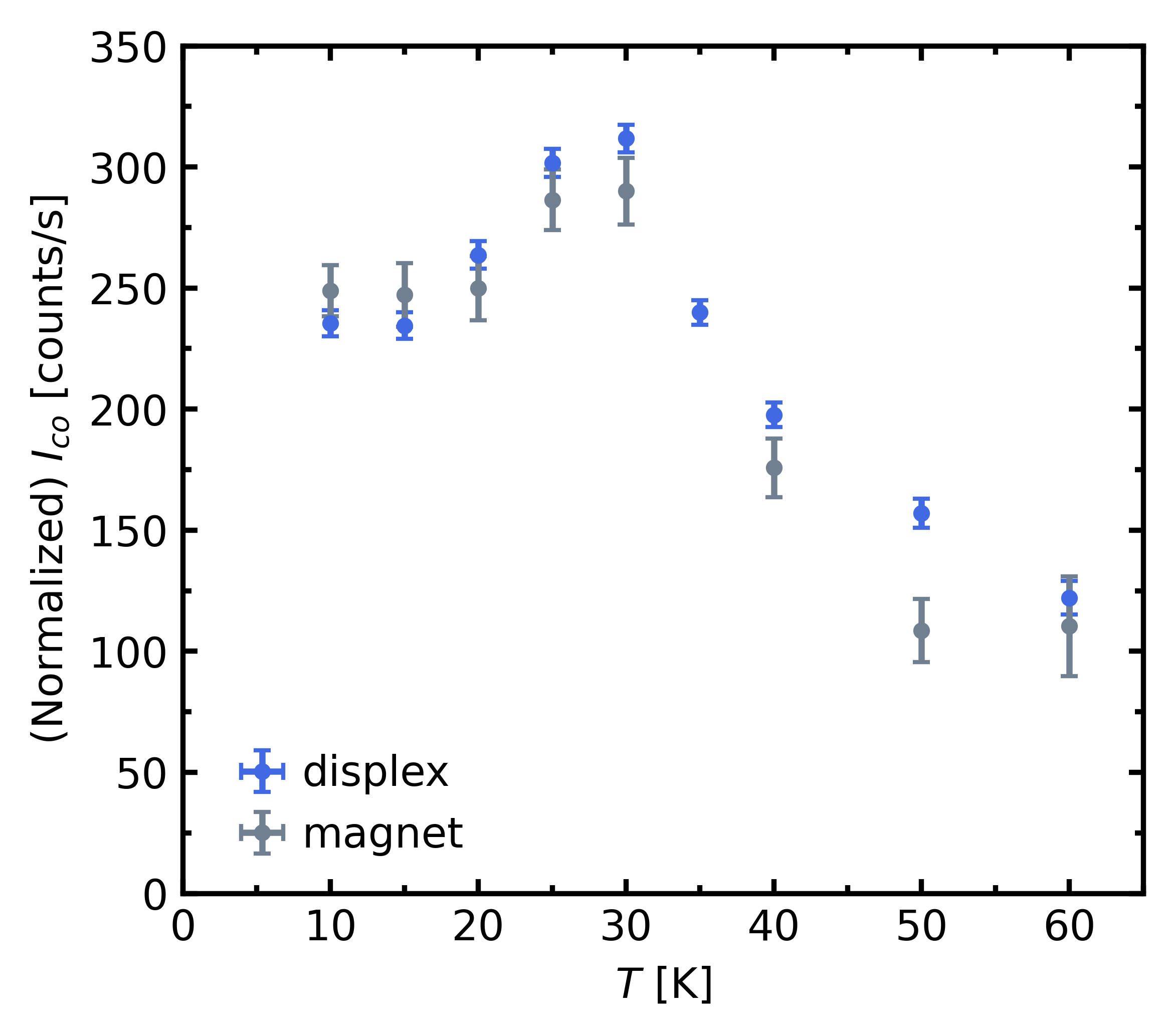}} 	
\caption{\textbf{Temperature dependence of charge order amplitude in different experimental environments.} Both data sets stem from the same unstrained LSCO sample. Blue data were measured in the displex on an Euler cradle, while grey data were measured in a cryomagnet. Error bars are standard deviations from fitting of charge order peak.}
\label{figs:compareexpenvir}
\end{figure}

\begin{figure}[h]
\center{\includegraphics[width=\textwidth]{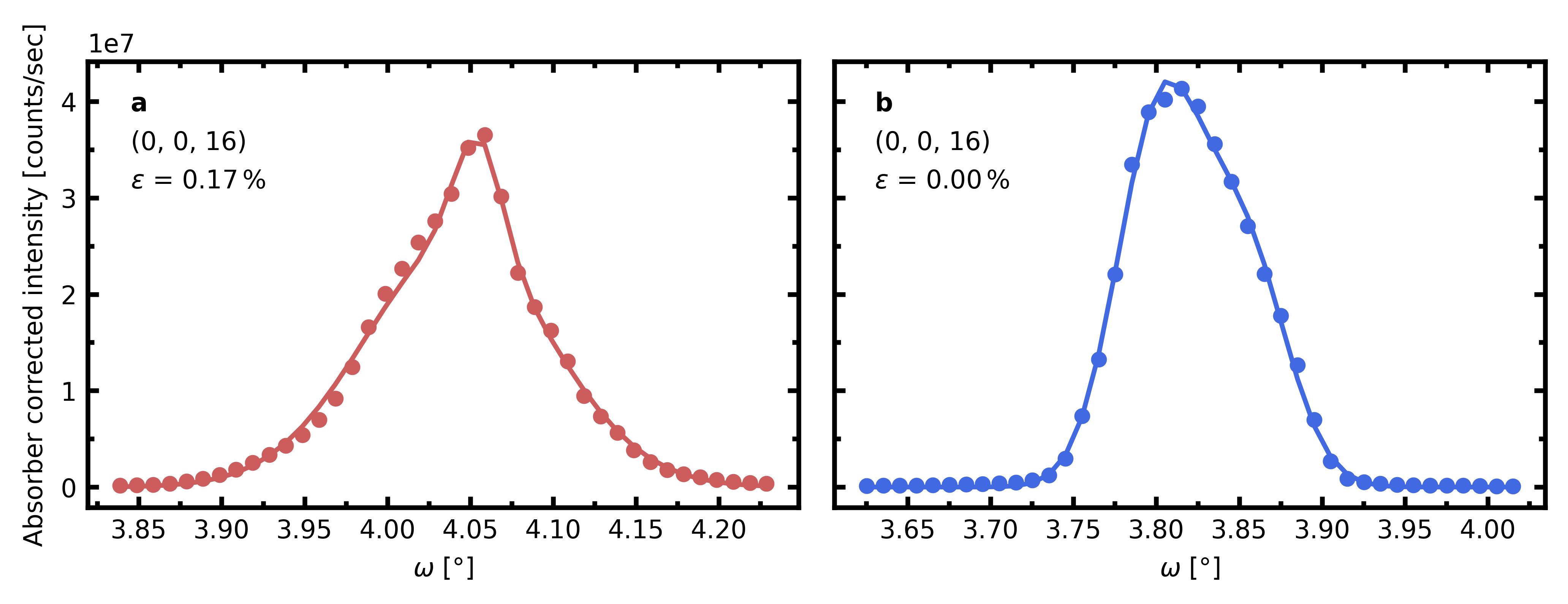}} 
\caption{\textbf{Comparison of Bragg peaks upon application of pressure.} \textbf{a} Bragg peak at $(0, 0, 16)$ with $\varepsilon_c = 0.17$\,\%.
\textbf{b} Same Bragg peak in ambient pressure conditions, after strain was released from the sample. Solid lines are guides to the eye. Error bars stem from counting statistics. Notice that errors are smaller than the point size and hence invisible.}
\label{figs:braggintensity}
\end{figure}

\begin{figure}[h]
\center{\includegraphics[width=0.5\textwidth]{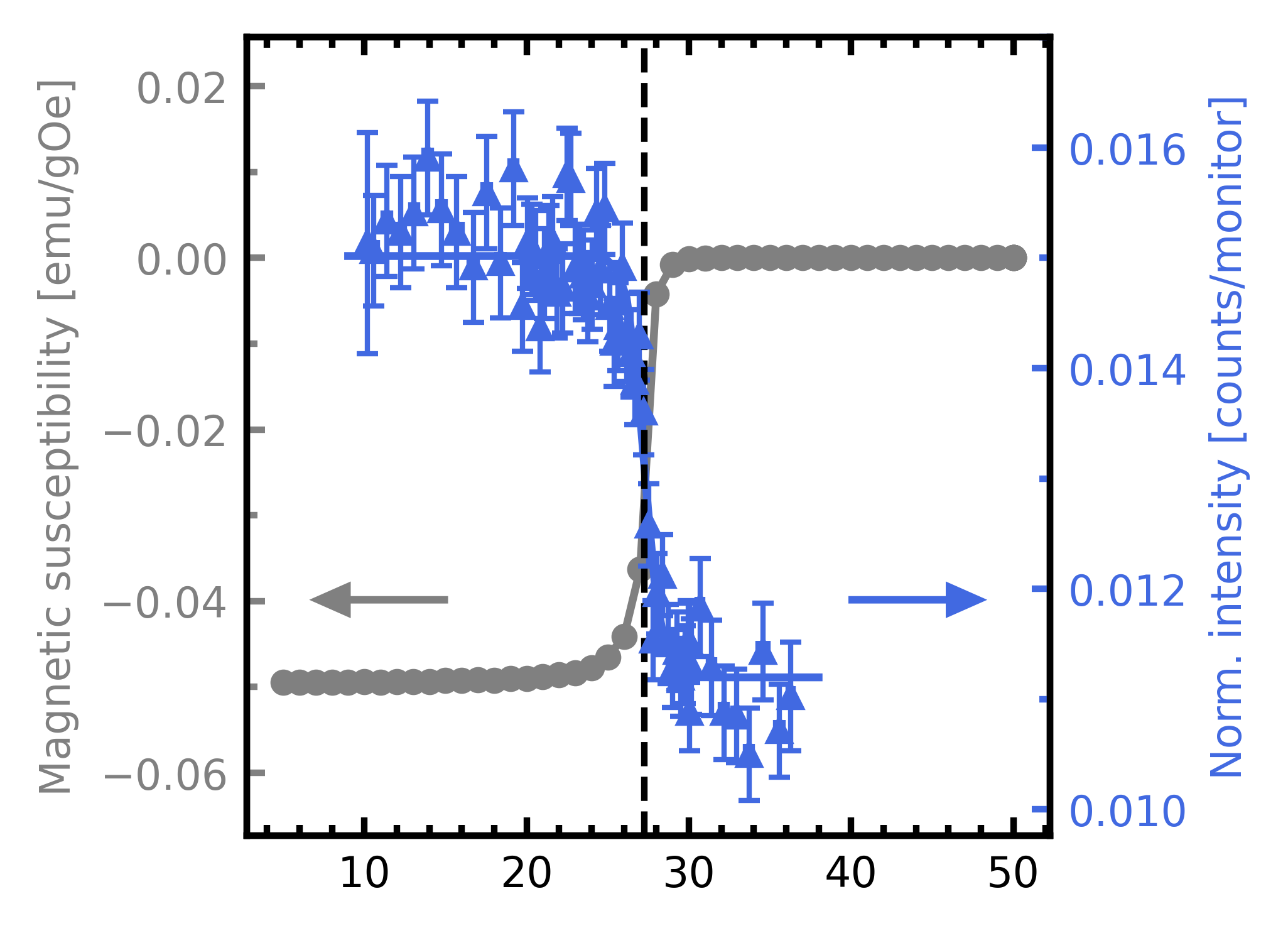}} 	
\caption{\textbf{Comparison of (zero-pressure) superconducting transition temperature} determined from zero-field-cooled DC magnetic susceptibility (grey, left) and polarized neutron scattering (blue, right). Error bars of neutron diffraction data represent counting statistics.}
\label{figs:tc}
\end{figure} 

\FloatBarrier

\begin{table}[h]
\vspace{3mm}
\begin{ruledtabular}
\begin{tabular}{ccccc}
Sample Environment & Crystal Batch & $(0, 0, n)$ & $(I^{0.00\,\%}_{int} – I^{0.17\,\%}_{int}) / (I^{0.00\,\%}_{int} + I^{0.17\,\%}_{int})$ &  $I^{0.00\,\%}_{int} / I^{0.17\,\%}_{int}$   \\ 
\hline
displex & I & $(0, 0, 4)$ & $0.126 \pm 0.006$ & $1.29 \pm 0.05$ \\
magnet cryostat & I & $(0, 0, 4)$ & $-0.041 \pm 0.003$ &  $0.92 \pm 0.06$\\
displex & I & $(0, 0, -12)$ & $-0.01 \pm 0.01$ & $ 1.0 \pm 0.1$ \\
displex & I & $(0, 0, 12)$ & $-0.01 \pm 0.01$ & $ 1.0 \pm 0.1$ \\
magnet cryostat & I & $(0, 0, -8)$ & $ -0.036 \pm 0.003$ &  $ 0.93 \pm 0.06$\\
magnet cryostat &  I & $(0, 0, 16)$ & $0.004 \pm 0.002$ & $1.01 \pm 0.06 $
\end{tabular}
\end{ruledtabular}
\caption{\textbf{Comparison of Bragg peak intensities.} Ratios $(I^{0.00\,\%}_{int} – I^{0.17\,\%}_{int}) / (I^{0.00\,\%}_{int} + I^{0.17\,\%}_{int})$ and  $I^{0.00\,\%}_{int} / I^{0.17\,\%}_{int}$ for the two different sample environments, including the Bragg peaks in Fig.\,\ref{figs:braggintensity}. 
$I_{int}$ is the integrated intensity and the superscript indicates the strain. All values bear upon $(0, 0, n)$ as indicated and the sample of batch I (see Tab.\,1). Integration is performed using the composite trapezoidal rule. Errors stem from biggest error estimation with the errors of the summed intensity of the ROI and Gaussian error propagation.}
\label{tab:tabs1}	
\end{table}

\section{Supplementary References}
\bibliography{Reference}